# Finding Alien Worlds in Queensland – A Decade of MINERVA-Australis

Jonathan Horner [1], Robert A. Wittenmyer [1], Stephen R. Kane [2], John Kielkopf [3], and Duncan Wright [1]

[1] *Centre for Astrophysics, University of Southern Queensland, West Street, Toowoomba, QLD 4350, Australia*
[2] *Department of Earth and Planetary Sciences, University of California, Riverside, CA 92521, USA*
[3] *Department of Physics and Astronomy, University of Louisville, Louisville, KY 40292, USA*

**Summary:** Three decades ago, humanity entered the Exoplanet Era, with the discovery of the first planets orbiting other stars. Today, more than 6000 exoplanets are known – a tally recently bolstered by NASA's *TESS* spacecraft. Whilst *TESS* is an exceptional planet finding machine, dedicated follow-up observations from the ground are required to confirm the existence of the planets it discovers. To achieve this, we constructed the southern hemisphere's only dedicated exoplanet detection and characterisation facility, MINERVA-Australis, at the University of Southern Queensland's Mt Kent Observatory. Funded in 2015, MINERVA-Australis saw first light in 2018, in time for the launch of *TESS*. MINERVA-Australis has since been scouring the skies, working to confirm and characterise the incredible harvest of planets detected by *TESS*. To date, the facility has contributed to the discovery of 40 new exoplanets, and continued the legacy of radial velocity data from the Anglo-Australian Planet Search program.

## Introduction

Over the past thirty years, we have lived through a great scientific revolution – the dawn of the Exoplanet Era. Where once we knew only a single planetary system – the Solar system – we now know planets are ubiquitous in the cosmos, and that almost every star is accompanied by a planetary retinue. This journey began in the 1980s, with the detection of the first debris discs, orbiting the young stars Fomalhaut, Beta Pictoris, and Vega [1][2][3], which provided evidence of debris left behind by planet formation. These discoveries were followed, in the early 1990s, by the detection of the first planets around other stars – the three planets (Draugr, Phobetor and Poltergeist) orbiting the millisecond pulsar PSR B1257 +12 (now named Lich) [4][5].

The true dawn of the Exoplanet Era is widely considered to have come with the discovery of the first planets orbiting Sun-like stars – 51 Pegasi b (Dimidium) [6], in 1995, which was quickly followed by 16 Cygni B b, 47 Uma b (Taphao Thong), 55 Cnc b (Galileo), 70 Vir b, τ-Boo b, and υ-And b (Saffar) [7][8][9][10]. For the first decade thereafter, planet discoveries came in slowly, with the overwhelming majority being made using the radial velocity (RV) method – using targeted observations of individual stars to measure changes in their radial motion over time by means of the Doppler effect[1].

The launch of the *Kepler* spacecraft, in 2009, lead to a significant uptick in the rate at which new exoplanets were discovered [12]. Taking advantage of the transit method of exoplanet detection, *Kepler* stared at a single patch of the sky for four years. It monitored the brightness of approximately 150,000 main sequence stars, looking for the periodic winks that suggest those stars are accompanied by planetary-mass companions – and proved hugely successful in that mission (e.g. [14][15]). To date, the data obtained by *Kepler* during its primary mission has led to the discovery of 2784 exoplanets[2]. The primary *Kepler* mission was brought to an end by the failure of two of the reaction wheels required to keep the telescope pointed at its target field – but the observatory was repurposed to carry out the K2 mission [16], which led to the discovery of a further 549 planets in the years that followed (e.g. [17][18]).

[1] For a detailed overview of the methods used to find exoplanets, we direct the interested reader to [11].

One of the core challenges presented by the data obtained by *Kepler* was that significant follow-up observations were required to confirm the existence of the potential planets identified by the spacecraft. Such observations were required to rule out and identify false-positive detections (e.g. [19][20][21]), and, for the planets that were confirmed, to provide additional characterisation (e.g. [22][23][24]). Many of the stars in the *Kepler* field were too faint for such observations to be easily made (e.g. [25]), and as such, there remain some 2955 'candidate exoplanets' detected by the *Kepler* spacecraft across its two missions that remain unconfirmed.

In 2018, NASA launched the successor to *Kepler*, the *Transiting Exoplanet Survey Satellite*, *TESS* [26]. Where *Kepler* was designed to stare at a single part of the sky, *TESS* is designed to carry out a survey of almost the entire night sky, searching for planets using the transit method. It observes strips of the sky that are 96° x 24° in size for a period of twenty-seven days, then rotates to observe an adjacent strip. In this manner, over the course of a single year, *TESS* can observe stars across an entire hemisphere of the night sky. As a result, *TESS* is delivering a constant stream of new exoplanet candidates that require follow-up observations from ground-based facilities to confirm and characterise [27]. At the time of writing, *TESS* has been involved in the discovery of 710 confirmed exoplanets and has generated thousands of additional planet candidates that require detailed additional observations to confirm[3].

In the years leading up to the launch of *TESS*, there was a growing awareness among the exoplanetary science community that there was a potential problem. *TESS* was expected to yield an incredible number of candidate planets that would require follow-up observations from ground-based facilities to confirm those discoveries, and to facilitate characterisation of the newly discovered worlds. As the New York Times put it, on the front page on Jan 6th, 2015 "So Many Earth-Like Planets, So Few Telescopes"[4]. The overwhelming majority of facilities used for exoplanetary science around the globe were, and still are, shared science instruments – where astronomers from all fields of study compete for time. As a result, there was significant concern in the community that the harvest delivered by *TESS* would be squandered, with candidate planets unable to be followed up and confirmed due to lack of global resources.

To address this problem, and to take advantage of wealth of data returned by *TESS*, we applied for funding from the Australian Research Council in 2015 to build a dedicated exoplanet follow-up and characterisation facility at UniSQ's Mt Kent Observatory. Using the $800,000 awarded from the ARC's LIEF scheme, along with funds from UniSQ and our partner universities, we were able to construct the MINERVA[5]-Australis facility [28][29] within just two years, and were on sky and gathering data by the time that *TESS* launched and saw first light. MINERVA-

[2] Data on the number of exoplanet discoveries throughout this work is taken from the NASA Exoplanet Archive [13], https://exoplanetarchive.ipac.caltech.edu/docs/counts_detail.html, accessed on 14th Nov 2025

[3] On 19th Nov 2025, the *TESS* Object of Interest (TOI) catalogue contained 7771 targets awaiting follow-up. The interested reader can access the TOI catalogue at: https://exofoipac.caltech.edu/tess/view_toi.php

[4] An archived online version of the article can be found at: https://www.nytimes.com/2015/01/07/science/space/as-ranks-of-goldilocks-planets-grow-astronomers-consider-whats-next.html

[5] MINERVA stands for 'Miniature Exoplanet Radial Velocity Array', named following our older partner observatory in the northern hemisphere (described in [33]).

Australis progressed from initial construction to science data acquisition faster than any other precise radial-velocity instrument on Earth.

In this work, to celebrate the tenth anniversary of funding for MINERVA-Australis, we present an update on the success of Australia's only dedicated exoplanet search and characterisation facility. We provide a description of the facility in the next section, before presenting an overview of the discoveries that have been made since MINERVA-Australis saw first light. We conclude with a discussion of the benefits of small, dedicated instruments like MINERVA-Australis, and look to the future to see what the next ten years hold for the facility.

## MINERVA-Australis

The MINERVA-Australis facility is located at the University of Southern Queensland's Mt Kent Observatory, approximately forty minutes south-southwest of the University's Toowoomba Campus. The observatory is located at 27.79738°S, 151.85587°E, at an elevation of 671 metres, under skies listed as Bortle class 2 (with an estimated zenith sky brightness SQM of 21.93 mag/arcsec$^2$), according to the data presented in the Light Pollution Atlas[6] [32]. Whilst the site does display some light pollution encroaching from Toowoomba (to the north northeast) and Brisbane and the Gold Coast (to the east), the skies from Mt Kent are exceptionally pristine for a site so close to major population centres.

MINERVA-Australis was built following the example set by the northern hemisphere MINERVA[7] facility – a dedicated exoplanet confirmation and characterisation facility located at Mt. Hopkins, Arizona [33][34]. The central concept is to have a facility that can carry out both radial velocity (RV) and transit observations using multiple telescopes to observe the stars thought to host exoplanets on every single clear night of the year. To achieve this, we took advantage of what can be considered the 'Ford Model T' revolution in telescope construction.

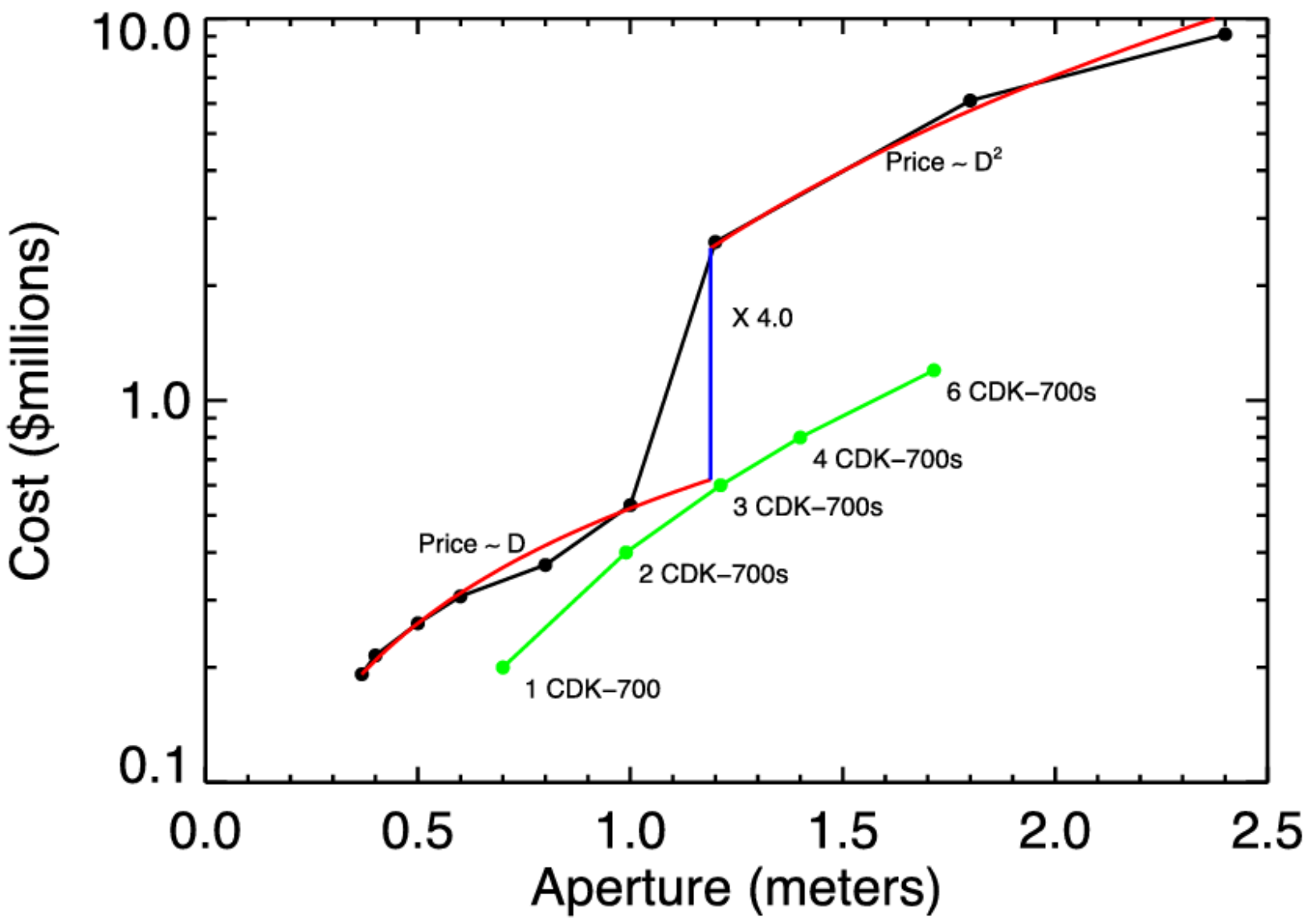

*Figure 1: MINERVA-Australis is an array of four PlaneWave Instruments CDK-700 telescopes, purchased off the shelf in 2017. Using research grade telescopes built on a production line, we obtain a collecting area equivalent to that of a 1.4m diameter custom-made research telescope, at a cost (shown in green) one order of magnitude cheaper than a bespoke instrument (whose price would lie on the red line). Figure reproduced from [33] and [28]; credit: Jason Wright.*

---

[6] Data taken from the Light Pollution Atlas website, https://www.lightpollutionmainfo/, on 9th Nov 2025

[7] Again, MINERVA is the Miniature Exoplanet Radial Velocity Array

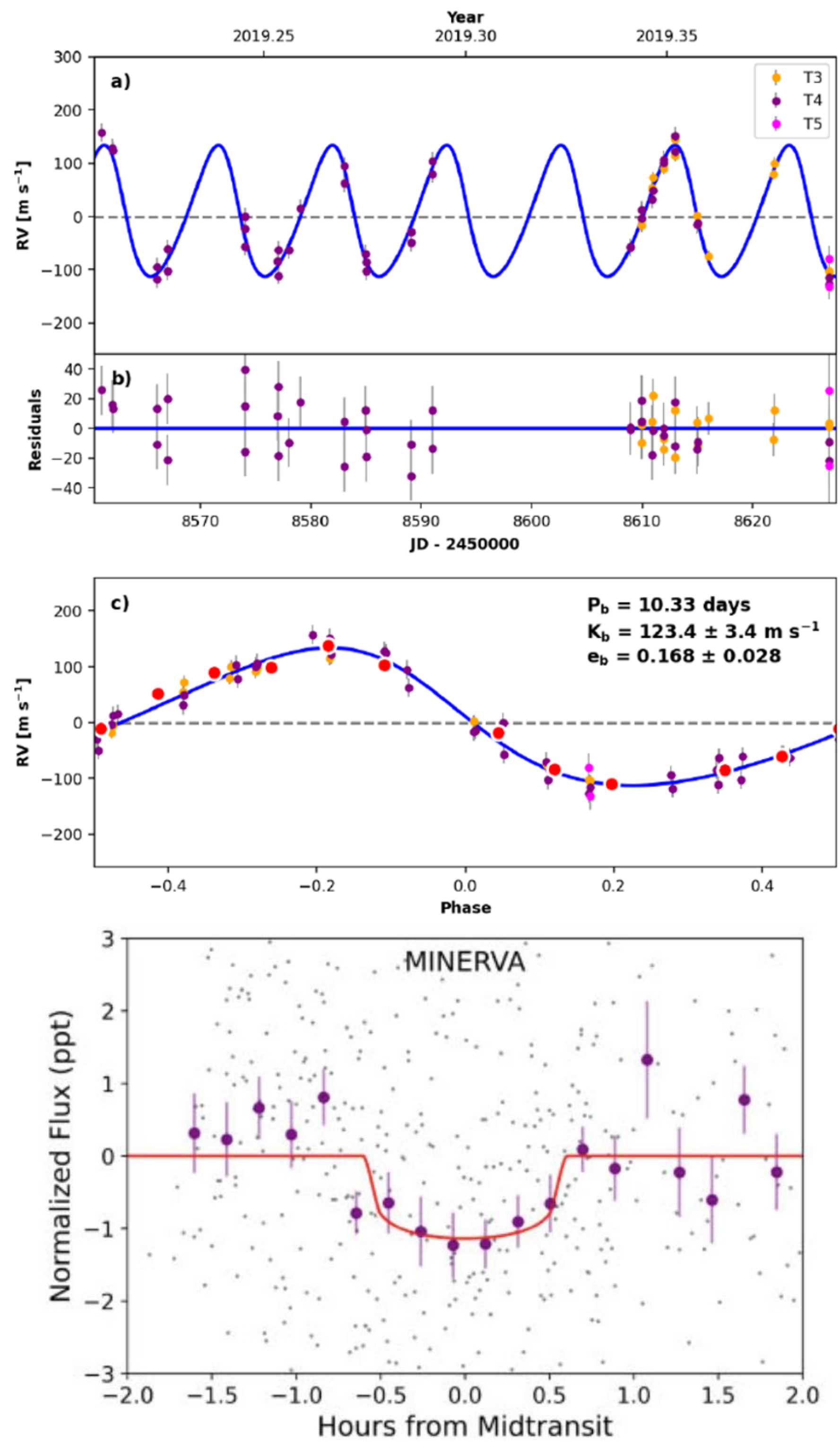

*Figure 2: The RV and photometric precision achieved by MINERVA-Australis. The upper two panels show the RVs obtained as part of the discovery of TOI-481b [30], whilst the lower panel shows the photometric observations we gathered as part of the discovery of Gliese 12b [31].*

Most research telescopes around the world are bespoke instruments. They are built for a specific purpose as a one off, and as a result, are extremely costly to build and maintain. Rather than designing bespoke telescopes for our facility, we instead purchased four PlaneWave CDK-700 telescopes – reflectors with a Corrected Dall-Kirkham optical design that are produced on a production line by PlaneWave Instruments in the United States of America. By purchasing telescopes that are built in bulk in this manner, we were able to achieve the effective light collecting area of a monolithic telescope with a mirror diameter of 1.4 metres at a cost approximately an order of magnitude lower than that we would have to pay for such a bespoke instrument (as shown in Figure 1, reprised from [28] and [33]).

By building a facility from several smaller telescopes rather than one large telescope, we were able to construct the southern hemisphere's only dedicated exoplanet characterisation and confirmation facility for a total cost of just a few million dollars. In addition, having multiple telescopes provides a level of flexibility that cannot be achieved from a single monolithic telescope. We can task different telescopes in the array to perform different kinds of analysis – for example, allowing simultaneous RV and transit photometry for a given target, or simultaneous broad band photometry in multiple filters at the same time.

The four CDK700 telescopes that make up the MINERVA-Australis array can each operate in photometric or RV mode, with the switch between modes carried out using the 45° tertiary flip mirror that directs light to either of the two Nasmyth focus ports on the altitude axis of the telescope mount. Whilst observing in RV mode, light gathered by the telescopes is fed through fibre optic cables to a high resolution (R~80,000) spectrograph in an insulated, environmentally controlled class 100,000 cleanroom in the main building. The spectrograph is held in a custom-made vacuum chamber that is thermally stabilised to ±0.01 K. The fibres that feed that spectrograph are aligned in the cross-dispersion direction of the spectrometer, forming four individual echelle traces. Two simultaneous fibres provide wavelength calibration and correct for instrumental variations. The calibration fibres are illuminated by a quartz lamp through an iodine cell, eliminating contamination by saturated Ar lines.

Details on the commissioning of MINERVA-Australis, along with the first photometric results, were presented in [29], and in the years since first light (achieved in 2018), the facility has consistently achieved a RV precision of ~10 m/s for *TESS* targets of V magnitude 10 or 11, with photometry for such stars consistently achieving a precision of a few millimagnitudes across a variety of optical and near-infrared broadband filters. Examples of the precision we can achieve in both RV and photometric observations are shown in Figure 2.

Mt Kent Observatory is located at a uniquely beneficial location in the southern hemisphere, allowing access to the south pole continuous viewing zone observed by *TESS*. Our position just outside the Tropic of Capricorn means that we are far enough north that we have access to a significant fraction of the northern sky – with MINERVA-Australis able to obtain useful observations of targets as far north as 40°N declination. Furthermore, our longitude means that we can observe targets when it is daylight is the Americas, Europe, and Africa – allowing us to obtain time-critical observations of events that would otherwise be inaccessible to the global exoplanet community. As a result, MINERVA-Australis has become a key component of the global effort to confirm and characterise the planets discovered by NASA's *TESS* spacecraft, to the extent that the US space agency purchased a significant fraction of the observing time on the facility for several years to guarantee access to observers from US institutions.

## A Plethora of Planet Discoveries

To date, MINERVA-Australis has been involved in the discovery of 40 exoplanets. In addition, we are currently performing our observations of several additional *TESS* Objects of Interest (TOIs), discoveries that will likely be announced in the coming years. A brief description of the exoplanet discoveries led by MINERVA-Australis is presented below, followed by a summary of the other discoveries to which our facility has contributed.

**TOI-257b – the First MINERVA-Australis led discovery**

The planet TOI-257b was the first discovery led by the MINERVA-Australis team [35]. Our team combined photometric observations made by the *TESS* spacecraft with RV measurements made using the MINERVA-Australis array, and data contributed by colleagues using the FEROS [36] and HARPS [37] facilities at La Silla Observatory. TOI-257 is a late F-type star that is somewhat evolved, with a mass 1.390±0.046 $M_{\odot}$, a radius of 1.888±0.033 $R_{\odot}$, and a surface temperature of 6095±89K. *TESS* photometry identified three transit events with a depth of ~1500 parts per million, with durations of approximately six hours. To confirm the existence of the transiting planet, and to help characterise it, we obtained 53 spectra of the star TOI-257 using MINERVA-Australis in the four-month period from July 12 to Oct 15, 2019. These data were supplemented by eight spectra obtained using FEROS (between 15$^{th}$ Dec 2018 and 22$^{nd}$ Jan 2019) and 33 spectra from HARPS obtained between Dec 2018 and Nov 2019.

Our analysis confirmed the existence of TOI-257b, a 'warm sub-Saturn', moving on an orbit with a period of approximately 18.38818±0.00085 days, at a semi-major axis of ~0.1528 au. The planet has a mass of 0.138±0.023 $M_J$[8], and a radius of 0.639±0.013 $R_J$. The bulk density inferred by this combination of radius and mass is ~ 0.650±0120 g cm$^{-3}$, almost identical to that of Saturn (687 kg m$^{-3}$). Given its proximity to the hot and luminous TOI-257, we estimate an equilibrium temperature for TOI-257b of approximately 1030 K.

In addition to the clear detection of the sub-Saturn TOI-257b, our data also contained evidence of a second signal, with a ~71-day orbital period. We are currently obtaining additional spectra of TOI-257 to determine whether that signal is real – so it is possible that we will be able to announce the discovery of a second planet in the system in the coming years.

**TOI-1842b**

The planet TOI-1842b is a particularly inflated (or 'puffy') object around two-thirds the mass of Saturn, orbiting a star markedly hotter and brighter than the Sun. That star, TOI-1842, has a mass 1.46 times that of the Sun, a radius of just over two Solar radii, and is currently approximately 5.5 times more luminous than our star. At approximately 2.49 Gyr old, the star has begun to evolve off the main sequence, puffing up and brightening as it moves towards the latter stages of its life. It was observed by *TESS* from 19 Mar to 15 Apr 2020, and that spacecraft witnessed two transit events, with a duration of ~4.4 hours and a depth of ~3150 parts per million, separated by a period of ~9.57 days.

To confirm the existence of TOI-1842b, photometric observations of three additional transits of the planet were performed using the Las Cumbres Observatory Global Telescope (LCOGT; [39]) on 30 May 2020, 13 Feb 2021 and 2 Apr 2021, confirming the signal was periodic, with transits occurring at the expected 9.57d period (equivalent to an orbital radius of ~0.10 au). The transit depths yielded a radius for TOI-1842b slightly larger than that of Jupiter (~1.04 $R_J$).

RV observations of TOI-1842b were made using MINERVA-Australis (119 spectra between 12 May 2020 and 3 Aug 2020), the Network of Robotic Echelle Spectrographs (NRES; [40]) on the LCOGT network (12 spectra between 7 June 2020 and 22 July 2020), and a single spectrum obtained on 27 May 2020 using the Fibre-fed Échelle Spectrograph (FIES; [41]). In combination, these spectra revealed that the planet is markedly less massive than Saturn, with a mass of ~0.214 $M_J$[9]. When this mass is taken together with the radius, as determined through our transit observations, they yield a bulk density for the planet of just ~0.252 g $cm^{-3}$ – placing it among the least dense planets discovered to date.

[8] In this context, $M_J$ denotes the mass of Jupiter (1.898×10$^{27}$ kg), and $R_J$ refers to the mean radius of Jupiter, 69,911 km. Similarly, $M_{\oplus}$ refers to the mass of the Earth (5.972×10$^{24}$ kg), and $R_{\oplus}$ is Earth's radius (6371 km).

TOI-1842b is a remarkable object. Its low density and high equilibrium temperature (1210K) tell the story of a planet reaching the end of its life. In the past, prior to TOI-1842 departing the main sequence, it seems likely that TOI-1842b was a fairly unremarkable sub-Saturn – but since the star has begun to accelerate away from the main sequence, its luminosity has risen markedly, and it is likely that TOI-1842b will soon begin to be whittled away by its increasingly luminous host. Eventually, the planet is fated to be devoured – in just a few hundred million years, TOI-1842 will expand to become a red giant, and TOI-1842b will be engulfed.

**TOI-778b**

TOI-778b is an ultra-hot hot Jupiter orbiting a young F3V dwarf star, HD115447 [42]. It was a particularly challenging discovery because of the rapid rotation rate of the star (v sin i = 35.1±1.0 km $s^{-1}$) – and was one of the first planets discovered by *TESS* orbiting a rapidly rotating star. *TESS* observed TOI-778 during Sector 10 of the primary mission (26 Mar to 21 Apr 2019), and again during Sector 37 (2 to 28 Apr 2021). During that time, it recorded eight transit events, with depths of approximately 8000 ppm. Additional photometric observations were carried out by the Next Generation Transit Survey (NGTS; [43]), which observed the planet's transit egress on 22 June 2019, and the Perth Exoplanet Survey Telescope (PEST)[10], which captured the egress of the planet during the transit on 31 Mar 2020. A full transit of the planet was observed photometrically using LCOGT [39] on 30 May 2020, with MINERVA-Australis being used in photometric mode to observe the full transit on 5 June 2020, and a final set of transit photometry was obtained from Mt Stuart Observatory in New Zealand, on 28 Apr 2020.

In addition to our photometric observations, we combined many RV measurements of TOI-778 from several facilities worldwide. Given the challenging nature of the host star (being both young and rapidly rotating), a larger number of RV observations were required than are typically needed for the confirmation of a massive transiting exoplanet. We therefore obtained 71 spectra using three of the four telescopes in the MINERVA-Australis array between 13 June 2019 and 4 June 2020; 13 spectra using the Tillinghast Reflector Echelle Spectrograph (TRES; [44]), between 15 June 2019 and 29 Feb 2020; 28 spectra using the CORALIE spectrograph [45], between 17 June and 1 Sept 2019; and 27 spectra using the CHIRON spectrograph [46], between 28 and 30 June 2019.

In combination, these varied observations revealed a picture of an incredibly hot planet almost three times the mass of Jupiter. TOI-778 b has a mass of ~2.76 $M_J$, a radius of ~1.37 $R_J$, and a bulk density of ~1.30 g $cm^{-3}$. It moves on an orbit with period just ~4.73 days, at a semi-major axis of just 0.06 au. The extreme insolation it receives from its host star leads to a calculated equilibrium temperature for the planet of 1561 K. Perhaps most interestingly, our results revealed that the planet, despite orbiting so close to its host, moves on a markedly eccentric orbit, with e = 0.21±0.04 that is tilted to the equatorial plane of TOI-778 by ~18±11°. Since the timescale for a planetary orbit so close to its host to become tidally circularised is very short, the combination of the planet's inclined and eccentric orbit suggests that either it is still undergoing a process of tidal circularisation, having been perturbed from a more distant orbit to a highly eccentric one with pericentre at the location of the current orbit – or that the planet is suffering from significant perturbations from other, as yet undiscovered, objects in the same system. As such, the TOI-778 system remains a fascinating target for future study.

[9] Radial velocity observations of a given star reveal the scale of the line-of-sight wobble of the star as it orbits around the barycentre of its mutual orbit with an unseen planet. Given knowledge of the mass of the star, this yields a minimum mass for the planet – in other words, the mass that the planet would need to induce such a wobble if its orbit were orientated perfectly edge-on to our line of sight. In the case of planets that are known to transit their host star (such as TOI-1842 b), the simple fact of the transits allows us to constrain the inclination of the planet's orbit with respect to our line of sight, confirming that the minimum mass derived from radial velocity observations is the true planetary mass (modulo the precision with which the host star's mass is known). For more detail, we direct the interested reader to [11], and references therein.

[10] The Perth Exoplanet Survey Telescope is run by amateur astronomer Thiam-Guan Tan; more details about the telescope and the exoplanet discoveries to which it has contributed can be found at https://pestobservatory.com/

**HIP 94235 A b**

HIP 94235 A b is a mini-Neptune (smaller in diameter than Neptune, but larger than the Earth), with a radius of ~3 $R_\oplus$, orbiting a young Sun-like G-type star (HIP 94235 A) with an orbital period of 7.7 days. [47]. That star is the more massive member of a binary star system, with the M-dwarf HIP 94235 B orbiting ~50 au from the star-planet pair. *TESS* observed the HIP 94235 binary star system during Sector 27 of its ongoing survey (4 – 30 July 2020), observing three small transit events during that time. Follow-up observations of the binary were carried out by the European Space Agency's *CHEOPS* satellite (the *Characterising ExOPlanets Satellite*; [48]), observing a transit of HIP 94235 b on 20 Aug 2021 – a single observation that was crucial in refining the ephemeris for the planet so that ground-based observations could be used for further study. Our team then obtained photometric observations from LCOGT to confirm no nearby stars could be contaminating the data, allowing us to confirm that the observed transits were not the result of interference from a background eclipsing binary star.

The spectroscopic observations of we obtained HIP 94235 A were unable to detect any RV signal from the planet. That was to be expected, as the mass of the planet is so small as to render it undetectable using the RV technique – particularly given the youth and high activity of the host star. Three spectroscopic observations were taken using the High-Resolution Spectrograph [49] on the South African Large Telescope [50], which allowed us to confirm the system was not a blended eclipsing binary system masquerading as a transiting exoplanet. We then obtained 16 observations of HIP 94235 using CHIRON, and 20 spectra using MINERVA-Australis.

Taken in concert, our results confirmed the presence of HIP 94235 A b, and allowed us to better understand the nature of its host star (HIP 94235 A), and the orbit of the binary system. HIP 94235 is a member of the AB Doradus moving group, which is ~120 Myr old. The detection of planets around such a young Sun-like star is particularly challenging – as is the detection of planets in close binary star systems. As a result, HIP 94235 A b stands as one of the smallest planets yet discovered orbiting a young star, and the HIP 94235 system is one of the tightest binary star systems yet found to host a planet orbiting one of the components.

**HIP 133103b and c**

Like the planet orbiting HIP 94235 A, the two planets orbiting HIP 133103 were discovered photometrically using observations from *TESS* and *CHEOPS* [51]. Follow-up photometric and spectroscopic observations were carried out using the ground-based CHIRON and MINERVA-Australis facilities, allowing us to confirm the validate the planetary nature of the transit candidates, and to characterise the properties of the host star, HIP 133103.

Our results confirmed the existence of the two planets, HIP 133103 b and c, moving on orbits close to mutual 2:1 mean-motion resonance, with periods of 7.610303 and 14.245651 days, respectively. HIP 113103 b is the smallest exoplanet yet confirmed in a MINERVA-Australis-led publication, with a radius of just ~1.83 $R_\oplus$, whilst HIP 133103 c is only slightly larger, at ~2.40 $R_\oplus$. Given the youth of the planet's host star (a K-dwarf with an estimated age of 470 Myr), it is quite likely that the two close-in planets are undergoing significant atmospheric loss – and this, coupled with the brightness of the host star, makes the system one of the most promising targets for future follow-up observations to study the formation and evolution of sub-Neptune planets orbiting close to their host stars.

**But wait, there's more…**

In addition to the MINERVA-Australis led discoveries described above, our facility has proved pivotal in the discovery of a significant number of other planets. Some highlights include:

- The three super-Earth planets orbiting the naked-eye F-type star HR 858 (HR 858 b, c, and d; [52]) – all of which have estimated equilibrium temperatures in excess of 1000 K, and orbital periods of just ~3.59, ~5.97 and ~11.23 days.
- The first planet found orbiting the second closest pre-main sequence star, AU Microscopii (AU Mic b; [53]) – a star with a spectacular spatially resolved edge-on debris disk. We built on that discovery by using MINERVA-Australis to measure the spin-orbit alignment for the planet using the Rossiter-McLaughlin effect ($\lambda = 47°^{+26}_{-54}$) – making it the youngest planet for which such a measurement has ever successfully been achieved [54].
- The three planets orbiting TOI-431 [55]– two super-Earths (TOI-431 b and c), and a sub-Neptune. Interestingly, in this system, only TOI-431 b and d transit their host star, with TOI-431 c only found through RV observations revealing a significant misalignment of the orbits of the three planets – something highly unusual for planets so close to their host star.

To date, our facility has been involved in the discovery of 40 planets. The great majority of these orbit stars that were identified as potential planet hosts by the *TESS* mission (as discussed above). However, we are not limited to solely performing follow-up for *TESS*.

## Added Value and Bonus Science

In addition to our core business providing follow-up observations of *TESS* candidate planets, we have also made extensive use of MINERVA-Australis to continue the legacy left behind by the Anglo-Australian Planet Search (AAPS; [56]). That program, which carried out RV observations using the 3.9m Anglo-Australian Telescope, came to an end in 2016, after discovering more than thirty exoplanets (e.g. [57][58][59] [60]). We are actively following up more than a dozen targets that were observed by the AAPS from 1998 to 2016, as part of our AAPS-Legacy survey. Such long-period RV surveys are currently the only reliable means by which we can find Jupiter-analog planets – planets of comparable mass to Jupiter moving on orbits with periods measured in decades (e.g. [61][62]).

To date, the AAPS-Legacy survey has only discovered a single planet – HD 83443 c [63]. HD 83443 c is a fascinating object – more massive than Jupiter ($m \sin i \sim 1.35\ M_J$), moving on an orbit with a semi-major axis of ~8.0 au (an orbital period of ~22 years), with an orbital eccentricity of ~0.76. That high eccentricity places it amongst the most eccentric exoplanet orbits known – with an orbit more reminiscent of the Solar system's Jupiter- and Saturn-family comets[11] than a typical planet! More interesting still is the fact that the HD 83443 system also contains a previously discovered hot Jupiter moving on a close, circular orbit around its host star. As such, the HD 83443 c system is a fascinating window into the chaotic dynamics that occur during the youth of planetary systems – with the most likely explanation for the extreme orbit of HD 83443 c being an ancient scattering event that flung that planet outward, whilst also slinging HD 83443 b inwards towards its host, making it the hot Jupiter we see today.

---

[11] For an extensive review of the Solar system in the context of Exoplanetary science, including a detailed discussion of the various families of cometary bodies therein, we direct the interested reader to [66].

Another example of the additional value we can extract by owning a dedicated RV observation machine is the recent work investigating the two short-period eclipsing binaries TIC 48227288 and TIC 339607421 [64]. In that work, we combined photometric observations carried out by *TESS* with RV measurements made by MINERVA-Australis to perform a highly detailed characterisation of those two close binary star systems – allowing us to measure the degree to which the orbital plane of those systems are aligned to the spin-axes of the stars therein. We found that both systems featured effectively circular orbits for the two component stars, but that those orbits were somewhat misaligned to the spin axes of the stars therein – a result that suggests current models of binary star formation and evolution are potentially incomplete.

In the years to come, we plan to expand our program of ancillary science with MINERVA-Australis – continuing the AAPS-Legacy survey, performing additional studies of newly discovered binary star systems, and performing observations of the Solar system's ice giant planets and small bodies. Our work on the AAPS-Legacy survey is a natural complement to the astrometric data expected to be delivered in the *GAIA* DR4 release – a set of data that will allow us to break the degeneracy between orbital inclination and planetary mass that means that RV observations alone can only obtain a minimum mass for the planets they reveal.

The primary focus of our facility will continue to be the follow-up and characterisation of candidate planets revealed by space-based transit surveys, such as *TESS*. At the time of writing, *TESS* has featured in the discovery of 710 confirmed exoplanets – but there are still 7771 candidate planets awaiting follow-up, according to the TOI catalogue. New candidates are added to that catalogue on a regular basis, meaning that it is highly unlikely that the MINERVA-Australis facility will run out of planetary candidates to study in the foreseeable future!

## Conclusions and Final Thoughts

The MINERVA-Australis facility, built at the University of Southern Queensland's Mt Kent Observatory in 2018, is Australia's only professional dedicated exoplanet discovery and follow-up facility. To date, it has been involved in the discovery of 40 confirmed exoplanets – many of which were first identified by observations made by NASA's *TESS* mission.

What is particularly remarkable about the success of the MINERVA-Australis facility is the incredible return on investment that we have achieved in the first seven years of observations. At a cost of just $6 million (AUD), we have built and operated a facility that has proven pivotal as part of the global effort to discover, confirm, and characterise alien worlds. Indeed, recent work ([65]) reveals that, when it comes to searching for Jupiter-mass planets, MINERVA-Australis is among the most cost-effective observatories on the planet.

It is a matter of great pride that these incredible results have been obtained by a relatively small facility in regional southeast Queensland – and we look forward to seeing what new surprises our facility uncovers during the next ten years of the MINERVA-Australis journey.

## Acknowledgements

We respectfully acknowledge the traditional custodians of all lands throughout Australia, and recognise their continued cultural and spiritual connection to the land, waterways, cosmos, and community. We pay our deepest respects to all Elders, ancestors and descendants of the Giabal, Jarowair, and Kambuwal nations, upon whose lands the MINERVA-Australis facility at Mt Kent observatory is situated. MINERVA-Australis is supported by Australian Research Council LIEF Grant LE160100001, Discovery Grants DP180100972 and DP220100365, the Mount Cuba Astronomical Foundation, and institutional partners the University of Southern Queensland,

UNSW Sydney, MIT, Nanjing University, George Mason University, the University of Louisville, University of California, Riverside, the University of Florida, and the University of Texas at Austin. This research has made use of the NASA Exoplanet Archive, which is operated by the California Institute of Technology, under contract with the National Aeronautics and Space Administration under the Exoplanet Exploration Program. The authors would like to express their sincere gratitude to the anonymous referee, whose feedback helped to improve the flow and readability of this review article.